\documentstyle[aps,prl,twocolumn]{revtex}
\tighten
\begin{document}
\draft
\title
{Strong-Electroweak Unification at About 4 TeV}
\author
{Paul H. Frampton $^{(1,2)}$}
\address{ $^{(1)}$ TH Division, CERN, CH1211 Geneva 23, Switzerland.}
\address
{$^{(2)}$ University of North Carolina,Chapel Hill, NC 27599, USA. }
\maketitle

\begin{abstract}
I show how an $SU(N)^{M}$ quiver 
gauge theory
can accommodate the standard model with three chiral families
and unify all of $SU(3)_C$, $SU(2)_L$ and $U(1)_Y$
couplings with high accuracy at one unique scale estimated as $M \simeq
4$ TeV.
\end{abstract}

\medskip
\bigskip
\medskip

Conformal invariance in two dimensions has had
great success in comparison to several condensed matter
systems. It is an
interesting question whether conformal symmetry
can have comparable success in a four-dimensional
description of high-energy physics.

Even before the standard model (SM) 
$SU(2) \times U(1)$ electroweak theory
was firmly established by experimental
data, proposals were made
\cite{PS,GG} of models which would subsume it into
a grand unified theory (GUT) including also the dynamics\cite{GQW} of
QCD. Although the prediction of
SU(5) in its minimal form for the proton lifetime 
has long ago been excluded, {\it ad hoc} variants thereof
\cite{FG} remain viable. 
Low-energy supersymmetry improves the accuracy of
unification of the three 321 couplings\cite{ADF,ADFFL}
and such theories encompass a ``desert'' between the weak
scale $\sim 250$ GeV and the much-higher GUT scale
$\sim 2 \times 10^{16}$ GeV, although minimal supersymmetric
$SU(5)$ is by now ruled out\cite{Murayama}.

Recent developments in string theory are suggestive
of a different strategy for unification of electroweak
theory with QCD. Both the desert and low-energy
supersymmetry are abandoned. Instead, the
standard $SU(3)_C \times SU(2)_L \times U(1)_Y$
gauge group is embedded in a semi-simple gauge
group such as $SU(3)^N$ as suggested by gauge
theories arising from compactification of the IIB superstring
on an orbifold $AdS_5 \times S^5/\Gamma$ where
$\Gamma$ is the abelian finite group $Z_N$\cite{F1}.
In such nonsupersymmetric quiver gauge theories
the unification of couplings happens not by
logarithmic evolution\cite{GQW} over an
enormous desert covering, say, a dozen orders
of magnitude in energy scale. Instead the
unification occurs abruptly at $\mu = M$ through the
diagonal embeddings of 321 in $SU(3)^N$\cite{F2}.
The key prediction of such unification shifts from
proton decay to additional particle content,
in the present model
at $\simeq 4$ TeV.

Let me consider first the electroweak group
which in the standard model is still un-unified
as $SU(2) \times U(1)$. In the 331-model\cite{PP,PF}
where this is extended to $SU(3) \times U(1)$
there appears a Landau pole at $M \simeq 4$ TeV
because that is the scale at which ${\rm sin}^2
\theta (\mu)$ slides to the value
${\rm sin}^2 (M) = 1/4$.
It is also the scale at which the custodial gauged
$SU(3)$ is broken in the framework
of \cite{DK}.

Such theories involve only electroweak
unification so to include QCD I examine the running
of all three of the SM couplings with $\mu$ as
explicated in {\it e.g.} \cite{ADFFL}.
Taking the values at the Z-pole 
$\alpha_Y(M_Z) = 0.0101, \alpha_2(M_Z) = 0.0338,
\alpha_3(M_Z) = 0.118\pm0.003$ (the errors in
$\alpha_Y(M_Z)$ and $\alpha_2(M_Z)$ are less than 1\%)
they are taken to run between $M_Z$ and $M$ according
to the SM equations
\begin{eqnarray}
\alpha^{-1}_Y(M) & = & (0.01014)^{-1} - (41/12 \pi) {\rm ln} (M/M_Z)
\nonumber \\
& = & 98.619 - 1.0876 y
\label{Yrun}
\end{eqnarray}
\begin{eqnarray}
\alpha^{-1}_2(M) & = & (0.0338)^{-1} + (19/12 \pi) {\rm ln} (M/M_Z)
\nonumber \\
& = & 29.586 + 0.504 y
\label{2run}
\end{eqnarray}
\begin{eqnarray}
\alpha^{-1}_3(M) & = & (0.118)^{-1} + (7/2 \pi) {\rm ln} (M/M_Z)
\nonumber \\
& = & 8.474 + 1.114 y
\label{3run}
\end{eqnarray}
where $y = {\rm log}(M/M_Z)$.

The scale at which ${\rm sin}^2 \theta(M) = \alpha_Y(M)/
(\alpha_2(M) + \alpha_Y(M))$ satisfies ${\rm sin}^2 \theta (M)
= 1/4$ is found from Eqs.(\ref{Yrun},\ref{2run})
to be $M \simeq 4$ TeV as stated in the introduction above.

I now focus on the ratio $R(M) \equiv \alpha_3(M)/\alpha_2(M)$
using Eqs.(\ref{2run},\ref{3run}). I find
that $R(M_Z) \simeq 3.5$ while $R(M_{3}) = 3$,
$R(M_{5/2}) = 5/2$ and
$R(M_2)=2$ correspond to
$M_3, M_{5/2}, M_2 \simeq 400 {\rm GeV}, ~~ 4 {\rm TeV}, {\rm and}~~ 140 {\rm TeV}$
respectively.
The proximity of $M_{5/2}$ and $M$, accurate to a few percent,
suggests strong-electroweak unification at $\simeq 4$ TeV.

There remains the question of embedding such unification 
in an $SU(3)^N$ of the type described in \cite{F1,F2}.
Since the required embedding of $SU(2)_L \times U(1)_Y$
into an $SU(3)$ necessitates $3\alpha_Y=\alpha_H$
the ratios of couplings at $\simeq 4$ TeV
is: $\alpha_{3C} : \alpha_{3W} :  \alpha_{3H} :: 5 : 2 : 2$ 
and it is natural
to examine $N=12$ with diagonal embeddings of
Color (C), Weak (W) and Hypercharge (H)
in $SU(3)^2, SU(3)^5, SU(3)^5$ respectively.

To accomplish this I specify the embedding
of $\Gamma = Z_{12}$ in the global $SU(4)$ R-parity
of the ${\cal N} = 4$ supersymmetry of the underlying theory.
Defining $\alpha = {\rm exp} ( 2\pi i / 12)$ this specification
can be made by ${\bf 4} \equiv (\alpha^{A_1}, \alpha^{A_2},
\alpha^{A_3}, \alpha^{A_4})$ with $\Sigma A_{\mu} = 0 ({\rm mod} 12)$
and all $A_{\mu} \not= 0$ so that all four supersymmetries are 
broken from ${\cal N} = 4$ to ${\cal N} = 0$.

Having specified $A_{\mu}$ I calculate the content
of complex scalars by investigating in $SU(4)$
the ${\bf 6} \equiv (\alpha^{a_1}, \alpha^{a_2}, \alpha^{a_3},
\alpha^{-a_3}, \alpha^{-a_2},\alpha^{-a_1})$ with
$a_1 = A_1 + A_2, a_2 = A_2 + A_3, a_3 = A_3 + A_1$ where
all quantities are defined (mod 12).

Finally I identify the nodes (as C, W or H)
on the dodecahedral quiver such that the complex scalars 
\begin{equation}
\Sigma_{i=1}^{i=3} \Sigma_{\alpha=1}^{\alpha=12}
\left( N_{\alpha}, \bar{N}_{\alpha \pm a_i} \right)
\label{scalars}
\end{equation}
are adequate to allow the required symmetry breaking to the
$SU(3)^3$ diagonal subgroup, and the chiral fermions
\begin{equation}
\Sigma_{\mu=1}^{\mu=4} \Sigma_{\alpha=1}^{\alpha=12}
\left( N_{\alpha}, \bar{N}_{\alpha + A_{\mu}} \right)
\label{fermions}
\end{equation}
can accommodate the three generations of quarks and leptons.

It is not trivial to accomplish all of these requirements
so let me demonstrate by an explicit example.

For the embedding I take $A_{\mu} = (1, 2, 3, 6)$ and for
the quiver nodes take the ordering:
\begin{equation}
- C - W - H - C - W^4 - H^4 -   
\label{quiver}
\end{equation}
with the two ends of (\ref{quiver}) identified.

The scalars follow from $a_i = (3, 4, 5)$
and the scalars in Eq.(\ref{scalars})
\begin{equation}
\Sigma_{i=1}^{i=3} \Sigma_{\alpha=1}^{\alpha=12}
\left( 3_{\alpha}, \bar{3}_{\alpha \pm a_i} \right)
\label{modelscalars}
\end{equation}
are sufficient to break to all diagonal subgroups as
\begin{equation}
SU(3)_C \times SU(3)_{W} \times SU(3)_{H}
\label{gaugegroup}
\end{equation}

The fermions follow from $A_{\mu}$ in Eq.(\ref{fermions}) as
\begin{equation}
\Sigma_{\mu=1}^{\mu=4} \Sigma_{\alpha=1}^{\alpha=12}
\left( 3_{\alpha}, \bar{3}_{\alpha + A_{\mu}} \right)
\label{modelfermions}
\end{equation}
and the particular dodecahedral quiver
in (\ref{quiver}) gives rise  to exactly {\it three}
chiral generations which transform under (\ref{gaugegroup})
as
\begin{equation}
3[ (3, \bar{3}, 1) + (\bar{3}, 1, 3) + (1, 3, \bar{3}) ]
\label{generations}
\end{equation}
I note that anomaly freedom of the underlying superstring
dictates that only the combination
of states in Eq.(\ref{generations})
can survive. Thus, it
is sufficient to examine one of the terms, say
$( 3, \bar{3}, 1)$. By drawing the quiver diagram
indicated by Eq.(\ref{quiver}) with the twelve nodes
on a ``clock-face'' and using
$A_{\mu} = (1, 2, 3, 6)$
in Eq.(\ref{fermions}) I find
five $(3, \bar{3}, 1)$'s and two $(\bar{3}, 3, 1)$'s
implying three chiral families as stated in Eq.(\ref{generations}).

After further symmetry breaking at scale $M$ to
$SU(3)_C \times SU(2)_L \times U(1)_Y$ the
surviving chiral fermions are the quarks and leptons
of the SM. The appearance 
of three families depends on both
the identification of modes in (\ref{quiver})
and on the embedding of $\Gamma \subset SU(4)$. The
embedding must simultaneously give adequate
scalars whose VEVs can break the symmetry 
spontaneously to (\ref{gaugegroup}).
All of this is achieved successfully by the
choices made.
The three gauge couplings evolve according to
Eqs.(\ref{Yrun},\ref{2run},\ref{3run}) for
$M_Z \leq \mu \leq M$. For $\mu \geq M$ the
(equal) gauge couplings of $SU(3)^{12}$
do not run if, as conjectured in \cite{F1,F2}
there is a conformal fixed point at $\mu = M$.

The basis of the conjecture in \cite{F1,F2}
is the proposed duality of Maldacena\cite{Maldacena}
which shows that in the $N \rightarrow \infty$
limit ${\cal N} = 4$ supersymmetric
$SU(N)$gauge  theory, as well as orbifolded versions with 
${\cal N} = 2,1$ and $0$\cite{bershadsky1,bershadsky2}
become conformally invariant.
It was known long ago
\cite{Mandelstam} that the
${\cal N} = 4$ theory is
conformally invariant for all finite $N \geq 2$.
This led to the conjecture in \cite{F1}
that the ${\cal N} = 0$
theories might be conformally
invariant, at least in some case(s),
for finite $N$.
It should be emphasized that this
conjecture cannot be checked
purely
within a perturbative framework\cite{FMink}.
I assume that the local $U(1)$'s 
which arise in this scenario
and which would lead to $U(N)$
gauge groups are non-dynamical, 
as suggested by Witten\cite{Witten},
leaving $SU(N)$'s.

This is a non-gravitational theory with conformal
invariance when $\mu > M$ and where the Planck mass
it taken to be infinitely
large. The ubiquitous question
is: What about gravity which breaks conformal
symmetry in the ultraviolet (UV)? This is
a question about the holographic principle
for flat spacetime.

From the phenomenological viewpoint the equal
couplings of $SU(3)^{12}$ can, instead of
remaining constant at energies $\mu > M$,
decrease smoothly by asymptotic freedom
to a conformal fixed point as $\mu \rightarrow \infty$.
This possibility is less restrictive and may fit
in better with the AdS/CFT
correspondence.
The desert 
resides in the unexplored domain of the
orders of magnitude in energy scale
between 4 TeV and the gravitational scale, $M_{Planck}$.

As for experimental tests of such
a TeV GUT, the situation at energies
below 4 TeV is predicted to be the standard model with
a Higgs boson still to be discovered at a mass
predicted by radiative corrections
\cite{PDG} to be below 267 GeV at 99\% confidence level.

There are many particles predicted 
at $\simeq 4$ TeV beyond those of
the minimal standard model.
They include 
as spin-0 scalars the states of Eq.(\ref{modelscalars}).
and 
as spin-1/2 fermions the states 
of Eq.(\ref{modelfermions}),
Also predicted are gauge bosons to fill out the gauge groups
of (\ref{gaugegroup}), and in the same energy region
the gauge bosons to fill out all of
$SU(3)^{12}$. All these extra particles are necessitated by
the conformality constraints of \cite{F1,F2} to lie
close to the conformal fixed point.

One important issue is whether this proliferation
of states at $\sim 4$ TeV is
compatible with precision
electroweak data in hand. This has
been studied in the related model of
\cite{DK} in a recent article\cite{Csaki}. Those results
are not easily translated to the present
model but it is possible that such an analysis 
including limits on flavor-changing neutral currents
could rule out the entire framework. 

As alternative to $SU(3)^{12}$ 
another approach to TeV unification has as its
group at $\sim 4$ TeV $SU(6)^3$ where 
one $SU(6)$ breaks diagonally to color
while the other two $SU(6)$'s each break to
$SU(3)_{k=5}$ where level
$k=5$ characterizes irregular embedding\cite{DM}.
The triangular quiver $-C - W - H - $
with ends identified and $A_{\mu} = (\alpha,
\alpha, \alpha, 1)$, $\alpha = {\rm exp} (2 \pi i / 3)$,
preserves ${\cal N} = 1$ supersymmetry.
I have chosen to describe the ${\cal N} = 0$
$SU(3)^{12}$ model in the text 
mainly because the symmetry breaking
to the standard model is 
more transparent.

The TeV unification fits ${\rm sin}^2\theta$
and $\alpha_3$, predicts three families,
and partially resolves the GUT hierarchy.  
If such unification holds in Nature there is a 
very rich level of physics one order of 
magnitude above presently accessible energy.

Is a hierarchy problem resolved in the present theory?
In the
non-gravitational limit $M_{Planck} \rightarrow \infty$
I have, above the weak scale, the new unification
scale $\sim 4$ TeV. Thus, although not totally resolved,
the GUT hierarchy is ameliorated.
The gravitational hierarchy problem is not addressed.

My final remark is on the non-appearance
of relevant deformations which break conformal invariance
above 4 TeV. This is an assumption I make
by analogy to several other systems in Nature
with a large scaling region, {\it e.g.} superfluid helium where 
there is a comparable non-appearance of relevant operators
over many orders of magnitude in scale size.

\bigskip
\bigskip

This work was supported in part by the
Office of High Energy, US Department
of Energy under Grant No. DE-FG02-97ER41036.

\end{document}